# Improvement of a focused ion beam fabricated diamond pillar for scanning ensemble nitrogen-vacancy magnetometry probe using an ultrapure diamond

Dwi Prananto[1*], Yifei Wang[1], Yuta Kainuma[1†], Kunitaka Hayashi[1], Masahiko Tomitori[1], and Toshu An[1‡]

[1]*School of Materials Science, Japan Advanced Institute of Science and Technology, Nomi, Ishikawa 923-1292, Japan*

[*]E-mail: prananto@jaist.ac.jp

[†]Present address: *Global Research and Development Center for Business by Quantum-AI technology (G-QuAT), National Institute of Advanced Industrial Science and Technology (AIST), Tsukuba, Ibaraki, 305-8560, Japan*

[‡]E-mail: toshuan@jaist.ac.jp

**Scanning diamond nitrogen-vacancy probe microscopy (SNVM) is an important tool for studying nanoscale condensed-matter phenomena. $Ga^+$-ion-focused-ion-beam (FIB) milling has been introduced as a method for fabricating SNVM, while the probe diameter is limited to a few micrometers due to $Ga^+$-induced damage. We report a method for improving the quality of FIB-fabricated SNVM probes by polyvinyl alcohol and Pt/Pd capping, followed by post-fabrication UV/ozone exposure. The effectiveness of the method is confirmed by the fabrication of an 800 nm-diameter probe with shallow NV centers, demonstrating an imaging of maze-like magnetic domain structure with a resolution of a few hundred nanometers and an improved sensitivity of 6.7 μT/Hz$^{1/2}$, while preserving spin coherence properties.**

Since its invention in 1982, scanning probe microscopy (SPM)[1] has played an indispensable role in studying physical and biological phenomena at the nanoscale[2]. The technique has been further developed to include nanoscale sensors attached to the tip apex of an SPM probe, enabling various functionalities and purposes. Among them, are the nitrogen-vacancy (NV) centers in diamond[3] (Fig. 1(a)) a multimodal quantum sensor based on optically addressable spin-triplet ($S = 1$) states[4], used for high-sensitivity and high-resolution sensing applications such as nanoscale magnetic domain structure imaging[5], nanoscale NMR[6], and temperature sensing of biological cells[7].

Recently, scanning diamond NV probe microscopy (SNVM) in the form of a diamond cantilever with a tip containing NV centers has been made possible by electron beam lithography (EBL) and reactive ion etching (RIE)[8], and is available commercially, e.g., Qnami AG and Qzabre AG[9]. $Ga^+$ ion-focused ion beam (FIB) has been used as an alternative method for diamond nanofabrication[10,11] and has recently been used to fabricate an SNVM probe from non-electronic-grade type-IIa diamond[12]. While offering greater 3D fabrication flexibility in shape and size, FIB-milling has been used to fabricate complex micrometer-sized photonic structures, such as a solid-immersion lens[11] and photonic crystal cavities[13]. However, FIB-milling suffers from $Ga^+$-ion-induced damage formation[14,15], limiting the fabrication of diamond structures to only a few μm, with NV centers located deep in the micrometer range beneath the surface[12].

To mitigate the FIB-induced damage on an FIB-fabricated SNVM probe, a soft inductively coupled plasma (ICP) etch has been demonstrated to effectively remove the damaged layer and improve NV spin coherence properties to some extent, while avoiding additional surface damage from the ICP etching[17]. Here, we introduce an alternative approach to preserve the near-surface ensemble of NV centers in an ultrapure diamond by capping with protective layers of polyvinyl alcohol (PVA) and Pt/Pd, followed by post-fabrication ultraviolet (UV)/ozone surface treatment to mitigate FIB-induced damage through surface chemistry modification without further damaging the diamond surface. This method enabled the successful fabrication of an SNVM probe with a sub-micrometer diameter while mitigating $Ga^+$-induced damage, thereby preserving the shallow NV ensemble. Moreover, evaluation of the fabricated SNVM probe indicates preserved and moderate improvement of spin-coherence properties, along with demonstrated nanoscale magnetic imaging and improved DC magnetic-field sensitivity over the previously reported FIB-fabricated SNVM probe[12].

An electronic grade (100)-oriented type-IIa diamond (Element Six) (measured $2 \times 2 \times 0.5$ mm$^3$), with initial impurity of N < 5 ppb and B < 1 ppb, was polished to be 50 μm in thickness. The diamond was then implanted with $^{14}N^+$ at the energy of 30 keV and a dosage of $1 \times 10^{12}$ ions/cm$^2$ and annealed at 900 °C for 1 hour under an Ar atmosphere to form a layer of NV centers at a mean depth of 40 nm (based on Monte Carlo simulation by SRIM[18]). Next, the diamond was acid-cleaned using a 1:3 mixture of $HNO_3$ and $H_2SO_4$ at 220˚C for 30 minutes and washed with deionized water. The diamond was then cut into triangular shapes by SYNTEK Co. Ltd using a laser beam. A triangular-shaped diamond was then picked up using a 30 μm diameter of electrochemically sharpened tungsten probe operated via a micromanipulator arm (Micro Support Co. Ltd.) and placed on an acetone-dissolved acrylonitrile butadiene styrene (ABS) substrate. The diamond was pushed against the dissolved ABS, acting as a binder for the micrometer-sized amorphous carbon residue formed on the diamond surface after the laser-cut process[19], thereby recovering a clean diamond surface (Fig. 2(a), upper left). The diamond was then transferred and fixed to a glass substrate using a fluoroelastomer adhesive (x-71-6046, Shin-Etsu) (Fig. 2(a) upper left).

$Ga^+$ ions FIB-milled on a diamond at 30 keV acceleration energy are known to create 30 nm-thick amorphous carbon residues and surface defects on the diamond[14], detrimental to the charge stability and coherence time of the NV centers[16]. Moreover, the FIB-milled diamond area is known to emit background fluorescence that may degrade the PL contrast of NV centers' optically detected magnetic resonance (ODMR)[11],[20],[21]. To protect the shallow NV centers from damage by $Ga^+$ ions and sample charging during the FIB milling process, the diamond was masked with a layer of polyvinyl alcohol (PVA) and subsequently with 160 nm Pt/Pd.

The SNVM probe was fabricated by a 30 keV $Ga^+$ ion beam in an FIB apparatus (SMI3050, SII NanoTechnology). The fabrication process was initialized by forming a square island of $7 \times 7 \times 5$ μm$^3$ using a rectangular milling pattern with 13 μA $Ga^+$ ion current to etch the area outside the square island. Next, a doughnut-shaped milling pattern of 7 μm outer diameter and 1 μm inner diameter was used to fabricate a 1.6 μm height pillar using a 720 pA $Ga^+$ ion current[12] (Fig. 2(a), upper-right panel). After fabrication, the diamond was soaked in a hot (60 °C) aqua regia solution (1:3 mixture of HCl: $HNO_3$) for 15 minutes to etch the Pt/Pd and PVA coatings, followed by 3 minutes of deionized water dipping (Fig. 2(a)). The SEM images in Fig. 1(c)-(d) show that the fabricated tip diameter was 800 nm.

After the fabrication process, we exposed the SNVM probe to UV/ozone to remove organic adsorbate and oxygen-terminate the surface[22)–27)]. The UV/ozone treatment was performed at room temperature and oxygen pressure of 0.1 MPa using a UV/ozone cleaner (UV253, Nippon Laser and Electronics Laboratory).

After the UV/ozone treatment, the diamond was mechanically detached from the glass substrate and picked up using the manipulator to be mounted at the apex of another 30 μm-diameter electrochemically sharpened tungsten rod attached to one prong of a quartz tuning fork (QTF) (Fig. 1(a)). The diamond was glued with silver paste (H20E, EPOTEK) and reinforced by epoxy glue (H74, EPOTEK). The SNVM probe was evaluated using a home-built scanning confocal microscopy (SCM) system consisting of a 532 nm laser source (MGL-III-532-80mW, Changchun New Industries Optoelectronics Technologies Co., Ltd.) guided through a 2D Galvano mirror system (GVS002, Thorlabs) to an objective lens (M Plan Apo 100× NA = 0.7, Mitutoyo) and focused on the tip of the SNVM probe. The laser polarized the NV center spins to the $m_s = 0$ state, emitting photoluminescence (PL) at 600–800 nm detected with an avalanche photodiode (SCPM-ARQH-14FC, Excelitas Technologies) through the same objective lens. A laser scan over an area around the SNVM probe reconstructed a PL image of the probe (Fig. 2(b)). Note that due to the 50 μm thickness of the current diamond pillar probe platform, no significant improvement in photon collection efficiency is observed compared to the EBL-RIE-fabricated SNVM probe with a thin (< 3μm) platform[28)]. However, further improvements are expected in the future by reducing the platform thickness to a few μm through additional FIB milling. A continuous-wave laser and microwave (SG6000F, DS Instruments) frequency sweep around 2.87 GHz resulted in an ODMR spectrum with a dip at 2.87 GHz at zero field (Fig. 2(c)). The PL Z-depth profile from the SNVM probe showed a full-width half-maximum of 2.3 μm, comparable to the 2.2 μm Z-axial diffraction limit of the objective lens (Fig. 2(d)), confirming the near-surface distribution of the NV centers in the ultrapure diamond as compared to the broadly distributed NV centers in our previously fabricated SNVM probe[12)].

The home-built SCM system was combined with a QTF-based phase-locked loop (PLL)-regulated atomic force microscopy (AFM) system to form an integrated SNVM system. The AFM system consisted of XYZ piezo translation motors (ANPx101 and ANPz101, Attocube Systems) for coarse probe movement and an XYZ piezo tube scanner for moving the sample relative to the probe. An FPGA system (USB-7845R, National Instruments) controlled the integrated AFM system. The tip of the diamond probe was brought into proximity to the sample by the Z-axis piezo tube scanner. The horizontally oscillating QTF was maintained in resonance by the PLL and provided an amplitude signal to feedback-control the tip-sample distance. The tip-sample distance was maintained, while the XY piezo tube scanner translated the sample to provide spatial information.

We evaluated the probe's spin coherence properties under a 10 mT external magnetic field aligned with one of the NV axes (<111> crystal-axis family). First, we checked the Rabi oscillation contrast ($m_s = 0 \leftrightarrow m_s = -1$) of the NV centers over an extended period of UV/ozone exposure. The Rabi oscillation was observed using a pulse protocol consisting of a 4 μs polarizing laser pulse (Pol.), followed by a resonant microwave pulse of varying duration and a subsequent 4 μs read-out laser pulse (RO1). An identical pulse sequence without MW provided a reference signal (RO2) for normalization (upper part of Fig. 3(a)).

The Rabi oscillation contrast, defined as the contrast between the maximum and minimum of the damped oscillation fit of the Rabi oscillation signals, was observed to be preserved at around 2 % (Fig. 3(b)). Spin coherence properties of a separate electronic grade diamond NV, prepared by the same implantation conditions of energy and dose and the post-NV generation surface acid treatment, were also measured and used as a control sample (referred as acid treatment dotted lines in Fig. 3). In comparison to the control sample, the Rabi contrast post UV/ozone treatment was observed to be improved by about 1.5 times. In all Rabi oscillation experiments, the microwave power was adjusted to produce a Rabi oscillation frequency of ≈ 1.3 MHz. The improvement in Rabi oscillation contrast is likely attributed to the increased oxygen-termination content on the diamond surface, predominantly in the form of C–O–H bonds, as studied by C. Li et al. using X-ray photoelectron spectroscopy (XPS)[27]. This surface oxygen termination suppresses upward band bending at the valence-band maximum of the diamond NV centers, as described by Yamano et al.[23].

The coherence times of the NV spins were inferred through the Hahn echo protocol with a MW sequence of $\pi/2 - \tau - \pi - \tau - \pi/2$ ($3\pi/2$) (upper part of Fig. 3(c)). The last MW pulse was altered between $\pi/2$ and $3\pi/2$ to rotate the superposition of $m_s = 0$ and $m_s = -1$ to respectively $m_s = 0$ (RO1) and $m_s = -1$ (RO2), to cancel common noise by subtracting RO1 and RO2. The longitudinal spin relaxation $T_1$ was evaluated by measuring the longitudinal spin relaxation of $m_s = 0$ and $m_s = -1$ (MW π-pulse) to their thermal equilibrium and subtracting the corresponding readout (RO1 and RO2, respectively) (upper part of Fig. 3(e)).

The coherence time $T_2$ Hahn echo of the fabricated SNVM probe was unchanged before and after the FIB fabrication process, followed by UV/ozone treatments, at the value of (30.4 ± 3.6) μs (Fig. 3(d)). This is reasonable, considering the $T_2$ Hahn echo is limited dominantly by the bulk paramagnetic substitutional nitrogen (P1 center) created after the ion implantation, and is less sensitive to the surface paramagnetic defect species[29)–31)].

On the other hand, the $T_1$ time was improved from (0.72 ± 0.015) ms before UV/ozone treatment to (1.83 ± 0.52) ms after 6 hours of UV/ozone treatment, comparable to the $T_1$ figure of (1.81 ± 0.31) ms of acid-treated diamond NV (Fig. 3(f)). $T_1$ improvement with UV/ozone exposure can be seen as the result of the improved surface state of the diamond, importantly around the rim of the diamond pillar, where C–O–H bonds predominantly terminate the surface, and the ratio of $sp^2$-bonded carbon (amorphous carbon) over $sp^3$-bonded carbon (diamond carbon) decreased[27)]. This improvement may indicate a reduction in FIB-induced damage states near the surface of the diamond, leading to reduced paramagnetic noise on the surface of the diamond[32)].

Finally, we demonstrated the applicability of the 800 nm-diameter SNVM probe for imaging the magnetic domain structure in a ferrimagnetic garnet, $(BiLu)_3Fe_5O_{12}$ (BiLuIG). BiLuIG, having in-plane magnetization and exhibiting a maze-like magnetic domain structure at zero field with Néel-type domain walls of a few hundred nanometers in width,[33)] is ideal for use as a standard test-bed material for evaluating our 800 nm-diameter SNVM probe. The sample was scanned over an area of 17.5 × 17.5 μm$^2$ (256 × 256 pixels) while pointing a continuous-wave laser on the NV centers probe and monitoring the PL. Fig. 4(a) shows the PL map showing the BiLuIG magnetic domain structure presented as bright areas with dark narrow lines structure. This all-optical magnetic domain imaging is enabled by stray magnetic fields with a high transverse component $B_\perp$ (>10 mT) generated by BiLuIG, resulting in mixed eigenstates of $m_s = 0$ and $m_s = \pm 1$ in both the ground and excited levels. Under laser excitation, this mixed state increased the decay rate and the probability of non-radiative transition from the excited state into the singlet state, thereby reducing NV spin polarization and quenching PL[34),35)]. In Fig. 4(a), dark lines indicate stray field vectors with a high off-axis component to the NV spins' quantization axis <111> that may be associated with domain wall structure (Fig. 4(b)), which had a width of about 300 nm inferred from the line cut of the periodic structure in the PL image (Fig. 4(c)). The domain wall width agrees with the estimated value of 400 nm based on the relation $\delta = \pi\sqrt{A/K_c}$, with $A = 3.7$ pJ/m as the exchange coupling constant and $K_c = 180$ J/m$^3$ as the cubic anisotropy of BiLuIG[33]. The estimated DC magnetic field sensitivity was determined from the slope of an ODMR spectrum of the fabricated probe. Due to the high transverse stray magnetic field, we were unable to obtain an ODMR spectrum in very close contact (a few nm) with BiLuIG; instead, we lifted the probe a few μm above the surface. The sensitivity, estimated as $\eta_{DC} \approx 1/\gamma_e(\partial I/\partial \nu)\sqrt{R}$, with $(\partial I/\partial \nu)$ is the steepest slope of the ODMR spectrum and $R$ (= 350 kcps) is the photon count-rate, was as small as 6.7 μT/Hz$^{1/2}$. The ODMR spectrum used to infer the DC magnetic field sensitivity is shown in Fig. 4(d). This value is better than the previously reported FIB-fabricated SNVM probe (12.3 μT/Hz$^{1/2}$)[12] and within the same order of magnitude as the reported sensitivities on the SNVM probes fabricated by EBL-RIE[8].

In summary, we fabricated an 800 nm-diameter SNVM probe using a laser cutting technique and Ga$^+$ ion FIB milling on an ultrapure diamond. We introduced a method for preserving the shallow NV center by PVA and Pt/Pd capping, followed by a room-temperature UV/ozone treatment, to mitigate the FIB-induced damage. Our SNVM probe preparation method preserved Rabi oscillation contrast and $T_2$, and recovers $T_1$ after a few hours of UV/ozone treatment. The SNVM probe was tested to image the magnetic domain structure in a ferrimagnetic BiLuIG and resolved the 300 nm magnetic domain wall structure with an estimated DC field sensitivity of 6.7 μT/Hz$^{1/2}$. It is worth noting that, due to the ensemble nature of NV in the SNVM probe, the performance of the FIB-fabricated SNVM scales down solely with the diameter of the pillar structure; with the present diameter of 800 nm, 25 NVs are estimated to be hosted in the pillar. The smaller the diameter, the better the spatial resolution at the cost of DC magnetic sensitivity due to the lower number of NVs[36]. In the future, it is worth investigating how the use of lower-acceleration energy (< 30 keV) of Ga$^+$-ion FIB milling can suppress the formation of the amorphous carbon layer and be used to fabricate probes with diameter < 200 nm with an estimated spatial resolution of about 75 nm, assuming a linear scaling-down of spatial resolution with probe diameter, for nanoscale condensed matter microscopy applications. Furthermore, the present work can be complemented by our previous work on ICP etching[17], in which ICP etching is used to remove FIB-induced damage and control the depth of the ensemble NV, followed by UV/ozone to chemically improve the surface condition.

### Acknowledgment

We thank N. Ito for the support in ion implantation, M. Ito for assisting with the FIB fabrication, and A. M. A. Hassan for the assistance in UV/ozone treatment. This work was supported, in part, by the JSPS KAKENHI Grant Nos. 24K01286 and 24K17580.

## Figures and captions

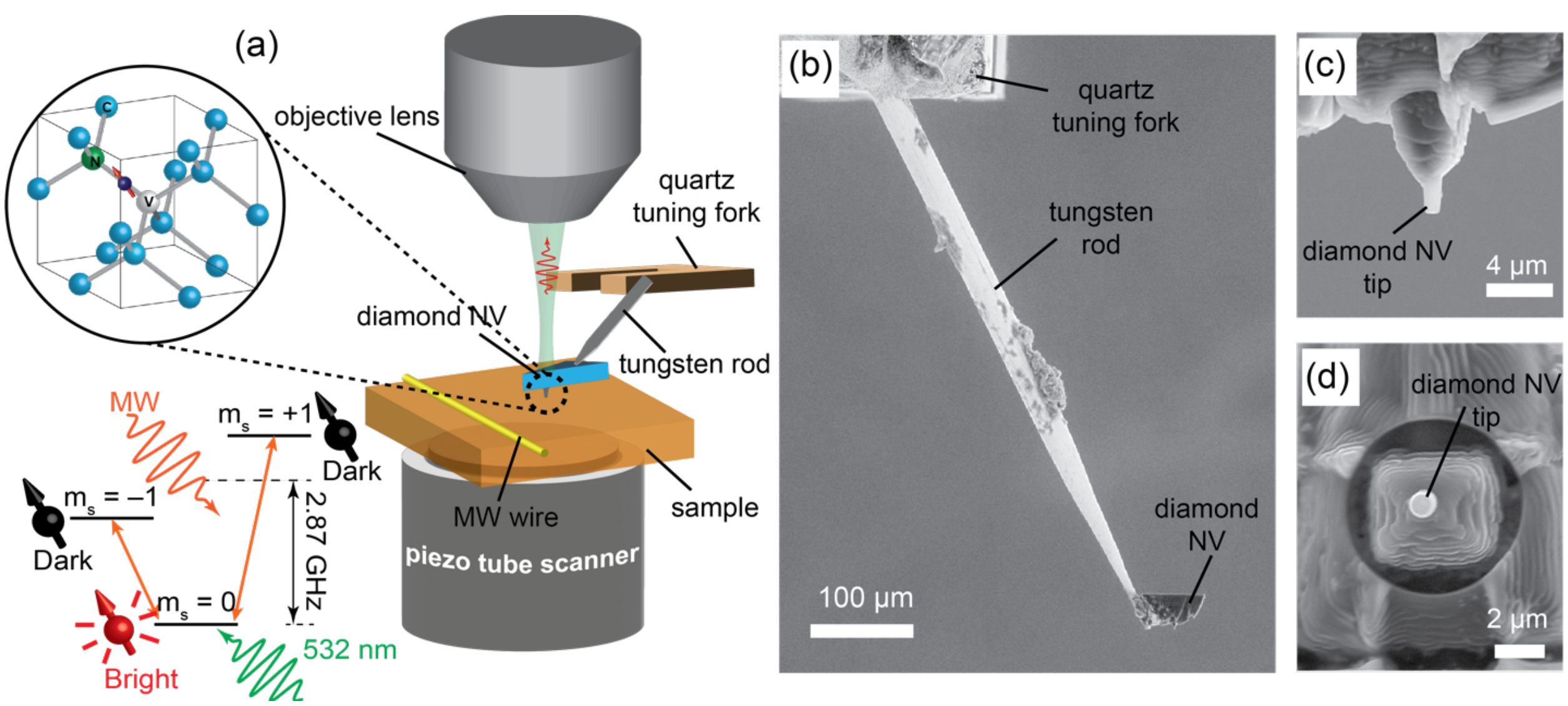

**Fig. 1.** (a) An FIB-fabricated SNVM probe was attached to the tip of a tungsten rod mounted on one prong of the quartz tuning fork, which was part of an integrated atomic force microscopy (AFM) system combined with a confocal microscopy system. (b) SEM image of the diamond NV probe attached to a tungsten rod. (c) Enlarged SEM image of the diamond NV probe showing the FIB-fabricated tip. (d) Same as (c) as seen from the bottom.

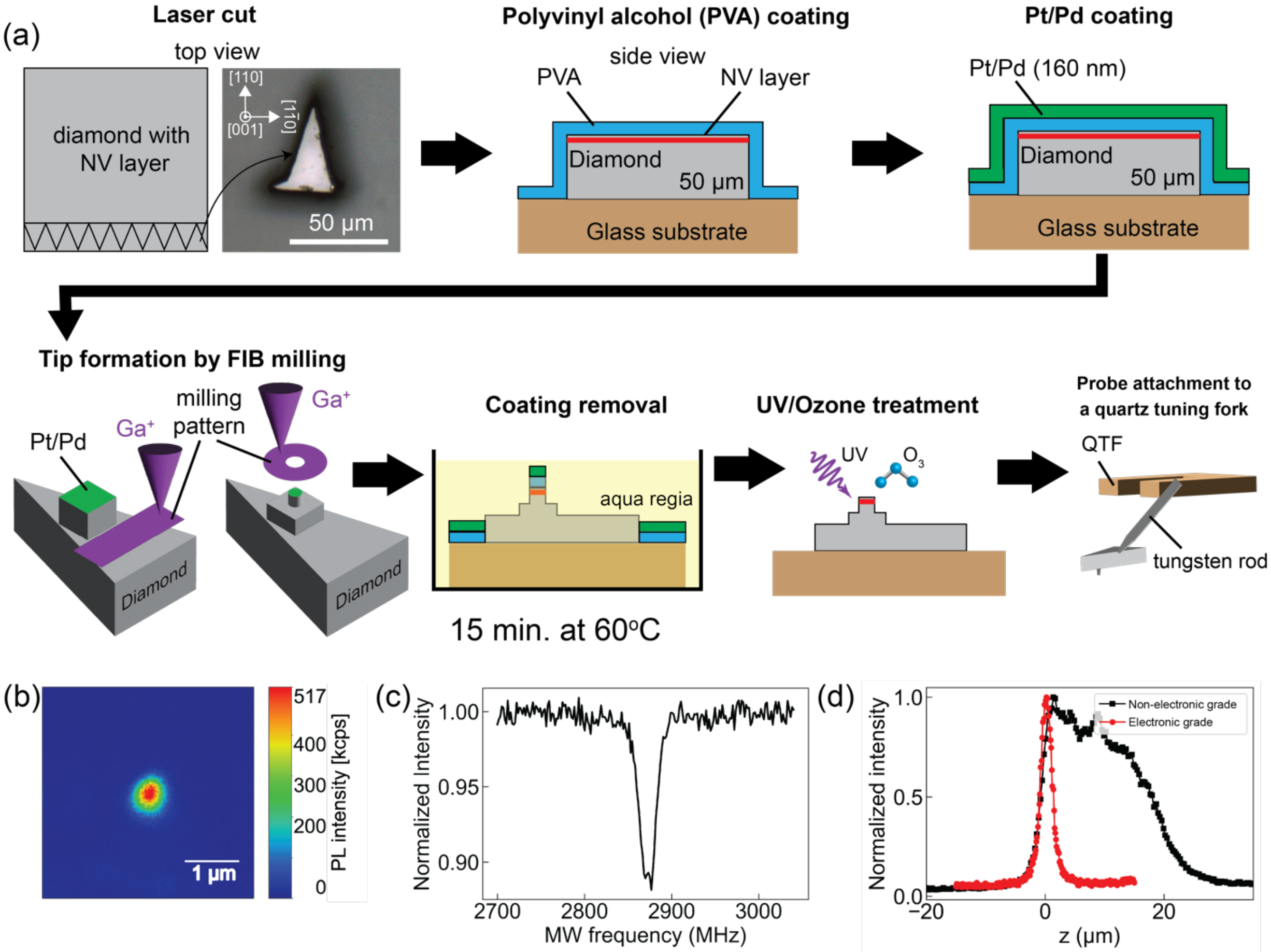

**Fig. 2.** (a) An SNVM probe fabrication process. A 50 μm thick diamond with an NV layer at 40 nm underneath its surface was cut into triangular shapes. A triangular-shaped diamond was picked up and masked with polyvinyl alcohol (PVA) and Pt/Pd (160 nm). The diamond was then milled with $Ga^+$ FIB to fabricate a pillar with a height of 1.6 μm and a diameter of 800 nm. The diamond was then dipped into a hot (60 °C) aqua regia for 15 minutes, followed by a DI water dip. Before being attached to a tungsten rod, the surface of the diamond probe was treated with UV/ozone exposure. (b) PL image of the SNVM tip. (c) The ODMR of the NV spins in the SNVM probe at zero external magnetic field, showing a drop in PL intensity at 2.87 GHz. (d) PL depth profile of NV diamonds produced in a non-electronic grade (black square) and an electronic grade (red circle) type IIa diamonds. FWHM of the electronic grade PL depth was 2.3 μm and 19 μm for the non-electronic grade diamond. Solid lines are a guide to the eyes.

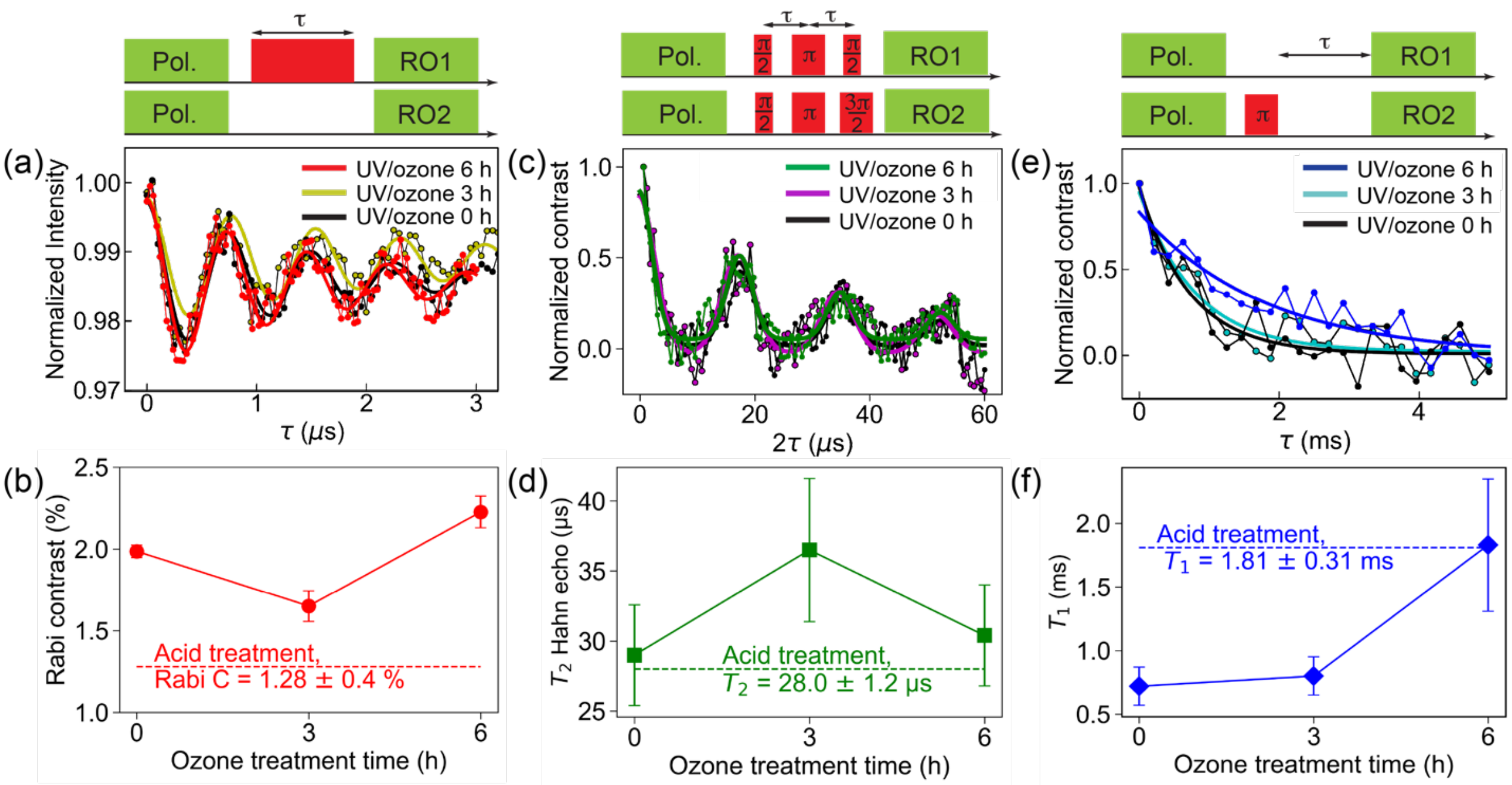

**Fig. 3.** (a) Rabi oscillation of the SNVM probe before (solid-black line), after 3 h (solid dark-yellow line), and after 6 h (solid-red line) of UV/ozone exposure. (b) Rabi oscillation contrast as a function of UV/ozone exposure time. (c) Hahn echo of the SNVM probe before (solid black line), after 3 h (solid-green line), and after 6 h (solid-green line) of UV/ozone exposure. (d) Hahn echo $T_2$ as a function of UV/ozone treatment time. The $T_2$ Hahn echo value before FIB fabrication (acid treatment) is shown as a dashed green line. (e) Longitudinal spin relaxation of the SNVM probe before (solid-black line), after 3 h (solid-cyan line), and after 6 h (solid-blue line) of UV/ozone exposure. (f) Longitudinal relaxation $T_1$ with increasing UV/ozone exposure time. The $T_1$ value before FIB fabrication (acid treatment) is shown as a solid blue line. Solid lines in (a), (c), and (e) are fitting lines, and solid lines in (b), (d), and (f) are a guide to the eyes. Error bars represent the standard deviation of the fitting to the experimental data.

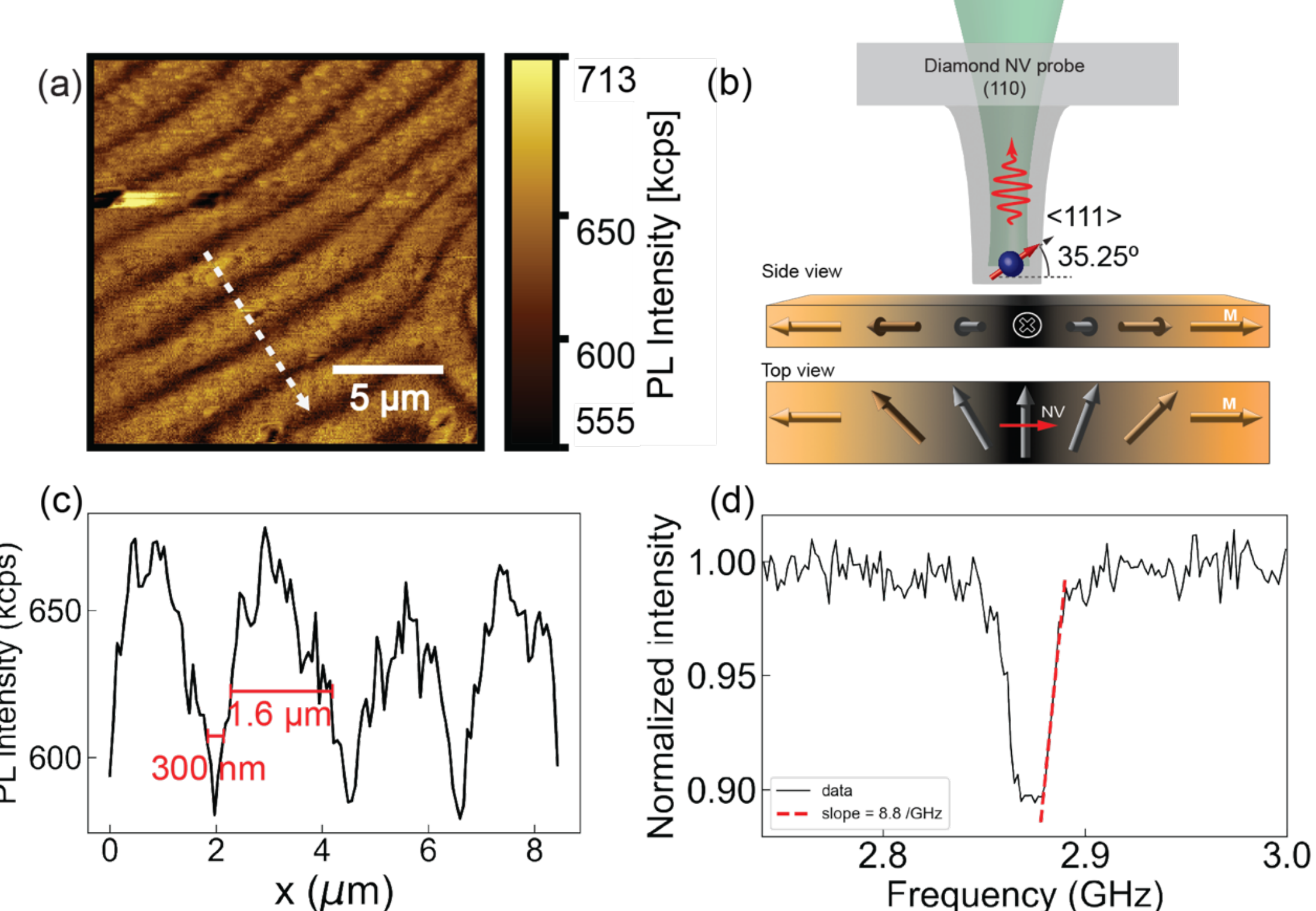

**Fig. 4.** (a) The diamond NV PL quenching image obtained as the probe was scanned over a $(BiLu)_3Fe_5O_{12}$. (b) Schematic illustration of the PL-quench imaging mechanism. The material's magnetic domain structure (pictured as in-plane domains with a Néel domain wall) was observed because the NV spins experienced strong and local off-axis stray fields, resulting in darker PL at positions with a high transverse stray magnetic field relative to the NV axis <111>. (c) Line-cut profile along the dashed line in (a). (d) An ODMR spectrum used to infer the DC-field sensitivity of the fabricated SNVM probe. The steepest slope is shown as a red dashed line.